# Detecting Land with Reflected-light Spectroscopy to Rule Out Waterworld $O_2$ Biosignature False Positives

Anna Grace Ulses[1,2,3], Joshua Krissansen-Totton[1,2,3], Tyler D. Robinson[4], Victoria Meadows[2,3,5], David C. Catling[1,2,3], and Jonathan J. Fortney[6]

[1] Department of Earth and Space Sciences, University of Washington, Seattle, WA, USA
[2] Astrobiology Program, University of Washington, Seattle, WA, USA
[3] NASA, Nexus for Exoplanet System Science, Virtual Planetary Laboratory Team, University of Washington, Seattle, WA, USA
[4] Lunar & Planetary Laboratory, University of Arizona, Tucson, AZ, USA
[5] Department of Astronomy, University of Washington, Seattle, WA, USA
[6] Department of Astronomy and Astrophysics, University of California, Santa Cruz, CA, USA

## Abstract

The search for life outside our solar system is at the forefront of modern astronomy, and telescopes such as the Habitable Worlds Observatory (HWO) are being designed to identify biosignatures. Molecular oxygen, $O_2$, is considered a promising indication of life, yet substantial abiotic $O_2$ may accumulate from $H_2O$ photolysis and hydrogen escape on a lifeless, fully (100%) ocean-covered terrestrial planet when surface $O_2$ sinks are suppressed. This so-called waterworld false-positive scenario could be ruled out with land detection because exposed land precludes extremely deep oceans (∼50 Earth oceans) given topographic limits set by the crushing strength of rocks. Land detection is possible because plausible geologic surfaces exhibit increasing reflectance with wavelength in the visible, whereas liquid water and ice/snow have flat or decreasing reflectance, respectively. Here, we present reflected-light retrievals to demonstrate that HWO could detect land on an exo-Earth in the disk-averaged spectrum. Given a signal-to-noise ratio (SNR) of 20 spectrum, Earth-like land fractions can be confidently detected with 0.3–1.1 $\mu$m spectral coverage (resolution $R \sim 140$ in the visible, $R \sim 7$ in the UV, with Earth-like atmosphere and clouds). We emphasize the need for UV spectroscopy down to at least 0.3 $\mu$m to break an $O_3$–land degeneracy. We find that the SNR and resolution requirements in the visible/UV imply a large aperture (∼8 m) will be necessary to ensure the observing times required for land detection are feasible for most HWO terrestrial habitable zone targets. These results strongly inform the HWO minimum requirements to corroborate possible oxygen biosignatures.

## 1. Introduction

The 2020 Astronomy & Astrophysics decadal survey prioritized a flagship mission specifically designed to search for life on exoplanets (National Academies of Sciences 2023). This Habitable Worlds Observatory (HWO) will survey ∼25 habitable zone exoplanets around Sun-like stars for biosignatures and habitability indicators in reflected light to understand how commonly life may evolve from habitable environments (National Academies of Sciences 2023). One potential barrier to the success of this mission is the presence of false-positive biosignatures—that is, nonbiological processes which mimic the impact of life on a planet's surface or atmosphere (S. D. Domagal-Goldman et al. 2014; V. S. Meadows 2017). This motivates early consideration of how to design HWO to successfully distinguish true biosignatures from false positives.

Molecular oxygen, $O_2$, is a widely studied biosignature due to its strong association with photosynthesis (T. W. Lyons et al. 2014), although planetary and stellar-driven processes are now known to also generate $O_2$, leading to false positives (V. S. Meadows et al. 2018). For example, abiotic $O_2$ may accumulate on planets orbiting M dwarf stars as a result of hydrogen loss during the star's superluminous pre-main-sequence phase (R. Luger & R. Barnes 2015) or the unique photochemistry generated by M dwarf hosts (S. D. Domagal-Goldman et al. 2014; P. Gao et al. 2015; V. S. Meadows 2017; C. E. Harman et al. 2018; S. Ranjan et al. 2023). In contrast, there are fewer well-studied abiotic $O_2$ accumulation mechanisms for more Sun-like F/G/K stars—which constitute the majority of suggested HWO targets (E. Mamajek & K. Stapelfeldt 2023). Nonetheless, modeling studies have identified scenarios wherein $O_2$-rich atmospheres can potentially develop on lifeless worlds orbiting Sun-like stars (R. Wordsworth & R. Pierrehumbert 2014; A. Kleinböhl et al. 2018; J. Krissansen-Totton et al. 2021).

In this paper, we focus on the feasibility of discriminating the waterworld $O_2$ false positive, which could be relevant for planets orbiting the F/G/K dwarfs that will dominate future HWO observations. A waterworld false positive may occur on rocky planets endowed with $H_2O$ inventories exceeding ∼50 Earth oceans (J. Krissansen-Totton et al. 2021). The pressure overburden from a large ocean increases the solidus of the silicate interior, rapidly stopping new crustal production (L. Noack et al. 2016; E. S. Kite & E. B. Ford 2018). The continuous production of fresh crust is necessary for long-lived geological sinks for $O_2$, such as degassing of reduced species and water–rock reactions (H. D. Holland 2002; D. C. Catling & J. F. Kasting 2017). Consequently, comparatively small $H_2O$ photolysis and H escape rates are sufficient to abiotically build up an Earth-like amount of $O_2$ in the atmosphere of a lifeless waterworld (J. Krissansen-Totton et al. 2021).

Water-rich terrestrial exoplanets (initial water inventory >50 Earth oceans) upon which abiotic $O_2$ may accumulate are expected to be abundant (S. N. Raymond et al. 2004, 2007; E. S. Kite & E. B. Ford 2018; T. Kimura & M. Ikoma 2022; R. Luque & E. Pallé 2022; A. R. Neil et al. 2022). In particular, S. N. Raymond et al. (2004) conducted 44 planet formation simulations, and found 41 of the resulting habitable zone planets had water contents exceeding Earth-like, 11 of which had >50 Earth oceans. There is also already direct evidence for the existence of water-rich exoplanets from JWST atmospheric characterization (C. Piaulet-Ghorayeb et al. 2024). Thus, the potential for oxygen false positives on waterworld exoplanets around Sun-like stars poses a problem for future life detection with HWO.

Fortunately, the waterworld false positive can be excluded by the detection of exposed land, which implies a limit on the ocean depth, since there is a predictable maximum height limit to topography based on the crushing strength of silicate rock (N. B. Cowan & D. S. Abbot 2014; C. M. Guimond et al. 2022). If HWO is capable of detecting land on an Earth-like exoplanet with an $O_2$-rich atmosphere, then it can be inferred that the ocean is not deep enough to provide the overburden pressure necessary to prevent new crustal production, and so there is a greater chance that any atmospheric $O_2$ is biogenic, or at least not abiotic through this particular mechanism.

Previous work has explored a variety of methods for detecting oceans and land on exoplanets, including spatial mapping using a planet's rotational light curve (N. B. Cowan et al. 2009), and ocean glint using phase-dependent and/or polarization observations (M. E. Zugger et al. 2010). Time-series observations of a rotating planet allow for spatial mapping of changes to the planet's atmosphere and surface (J. Lustig-Yaeger et al. 2018). Taken at multiple wavelengths, these observations can be used to infer a longitudinal map and color variations of surfaces on an exoplanet (N. B. Cowan et al. 2009). With further knowledge of the planet's obliquity and inclination, 2D color maps can be constructed and used to identify large exo-oceans and continents (H. Kawahara & Y. Fujii 2010, 2011). However, mapping oceans and continents via rotational variability is only possible for planets with large-scale longitudinal variations in albedo, therefore some exo-Earths would be unmappable due to unfavorable continent configurations. Furthermore, light-curve inversion techniques used to identify oceans and continents have typically required a priori assumptions about surface composition (N. B. Cowan et al. 2009; Y. Fujii et al. 2010).

Another powerful tool for identifying oceans is glint: the specular reflection attributable to flat or liquid surfaces (D. M. Williams & E. Gaidos 2008). When observed at crescent phase, glint increases planetary brightness, although this behavior can be mimicked to a lesser degree by forward-scattering by clouds (T. D. Robinson et al. 2010; D. J. Ryan & T. D. Robinson 2022). Observing a planet at crescent phase is not always possible due to unfavorable orbital inclination, and requires a small inner working angle that may not be possible for many HWO targets (S. R. Vaughan et al. 2023). A hybrid approach is to combine longitudinal mapping with phase-dependent glint measurements to identify oceans and continents (J. Lustig-Yaeger et al. 2018). However, this method requires high-cadence, high-SNR observations at crescent phase when the planetary flux is lower; it is estimated that only a single ocean may be detected this way for a 6 m class HWO telescope (J. Lustig-Yaeger et al. 2018).

Previous methods for detecting oceans and continents, while potentially powerful, may require inaccessible information on planetary inclination/obliquity, or small inner working angles and high-SNR observations at crescent phase where the planetary flux is low. Additionally, these methods typically do not consider the atmosphere as separate from the surface, and so any continent detection is based on a blended surface and atmosphere spectrum that may not necessarily be uniquely attributable to surface land. Here, we explore a new method for ruling out a waterworld false positive via direct land detection in the disk-averaged planetary spectrum using reflected-light retrievals. Ruling out a waterworld false positive in this way is predicated on the detection of a positive slope in the visible to near-infrared (NIR), which appears to be a ubiquitous feature of land surface spectra (see below, and Figure 1(a)). We present simulated reflected-light retrievals to show that a positively sloped land surface spectrum can be confidently inferred agnostically, even if atmospheric composition, cloud fraction, and bulk planetary properties are unknown.

## 2. Methods

### 2.1. Representative Surface Spectra

Unless stated otherwise, the nominal "true" planet for all retrievals is that of a simulated lifeless Earth-like planet with 60% ocean, 10% ice, and 30% land. "Lifeless" here means no biosphere is present to alter the atmosphere or surface. To broadly represent Earth's surface composition, this 30% land fraction is further divided into 5% weathered basalt, 5% kaolinite, 10% weathered granite, and 10% desert sand (M. Tang 2020). Kaolinite was chosen as a representative clay mineral produced by the weathering of silicates (L. N. Warr 2022). Our chosen desert fraction is based on the current areal extent of global deserts (J. Laity 2008). With the exception of the spectra of desert sand (G. Roccetti et al. 2024) and ice (J. Stapf et al. 2020), all spectra are taken from the USGS spectral library.[7] We adopted these representative surfaces in our retrievals of exo-Earths because (i) silicate surfaces are expected from the accretion and differentiation of Earth-sized rocky planets given cosmochemically abundant refractory materials (D. C. Rubie et al. 2011), and (ii) basaltic surfaces are common across our own solar system (L. T. Elkins-Tanton 2012). Furthermore, granitic and clay surfaces are expected due to ongoing crustal production on exo-Earths in the presence of surface water (K. H. Wedepohl 1995). Recently, JWST has provided compositional constraints for airless rocky exoplanets, suggesting that silicate surfaces are widespread (C. V. Morley et al. 2017; M. Gillon 2024; J. A. Patel et al. 2024). We account for the potentially greater diversity of exoplanet surface compositions later in this study in Section 2.4.

Figure 1(a) shows the reflectance spectra of all representative surfaces used in this study in the absence of an overlying atmosphere. The nominal lifeless Earth-like planet surface (no overlying atmosphere) is also shown, calculated from a weighted mixture of the other surfaces as described above. Our representative land surfaces all have positive slopes through the visible-NIR. This upward slope in the reflectance spectra is a general property of all silicate materials—we found no exception to this behavior in the USGS spectral library, or among any empirical solar system rocky surface spectra (J. H. Madden & L. Kaltenegger 2018). The explanation for this general property of rocky surface spectra has not been fully investigated, but it is seemingly attributable to electronic pair transition absorption in iron oxides (A. Valantinas et al. 2025). Crucially, it makes land observationally distinct from ocean/ice/snow surfaces, which are flat or have a negative slope, i.e., decreasing reflectivity with increasing wavelength.

[7] https://crustal.usgs.gov/speclab/QueryAll07a.php

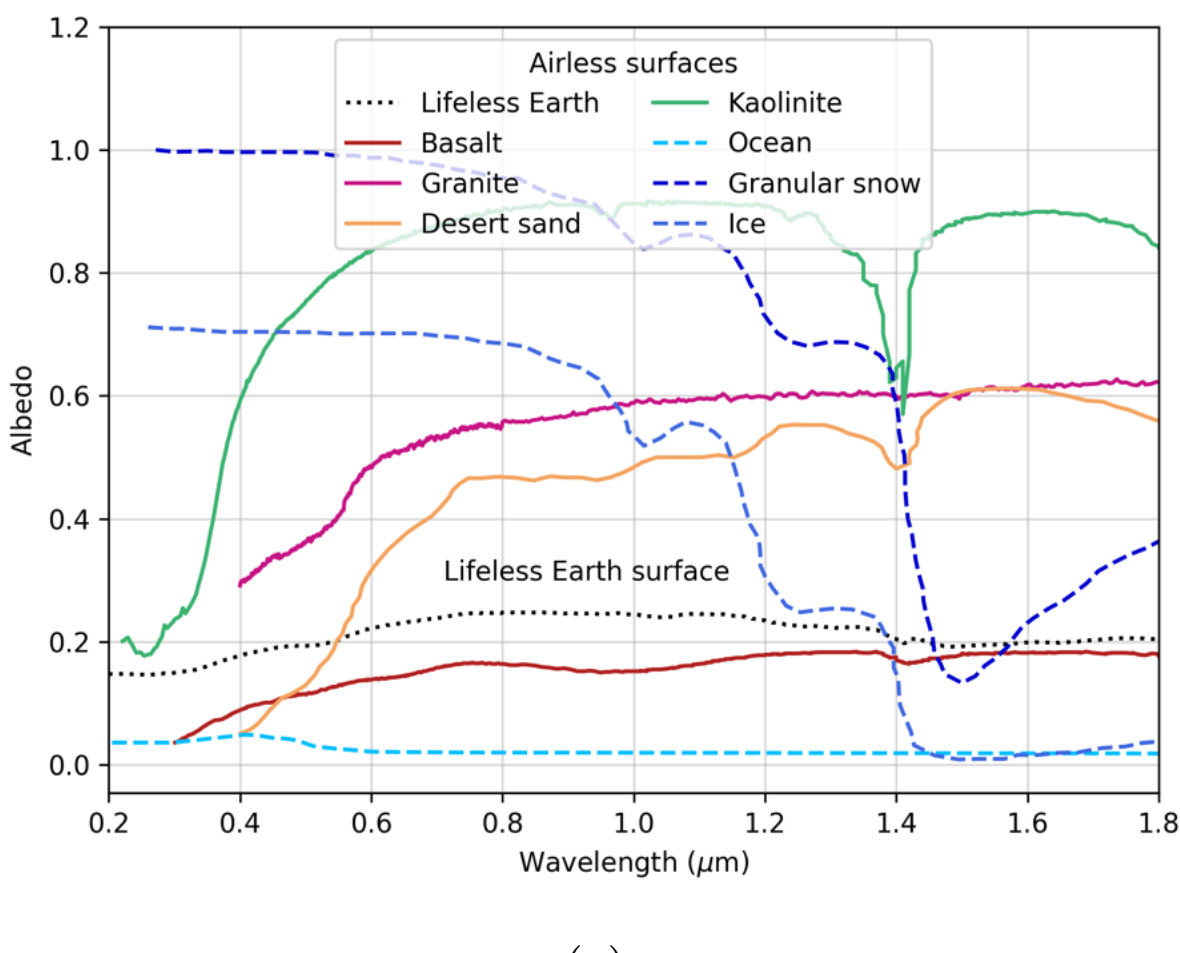

(a)

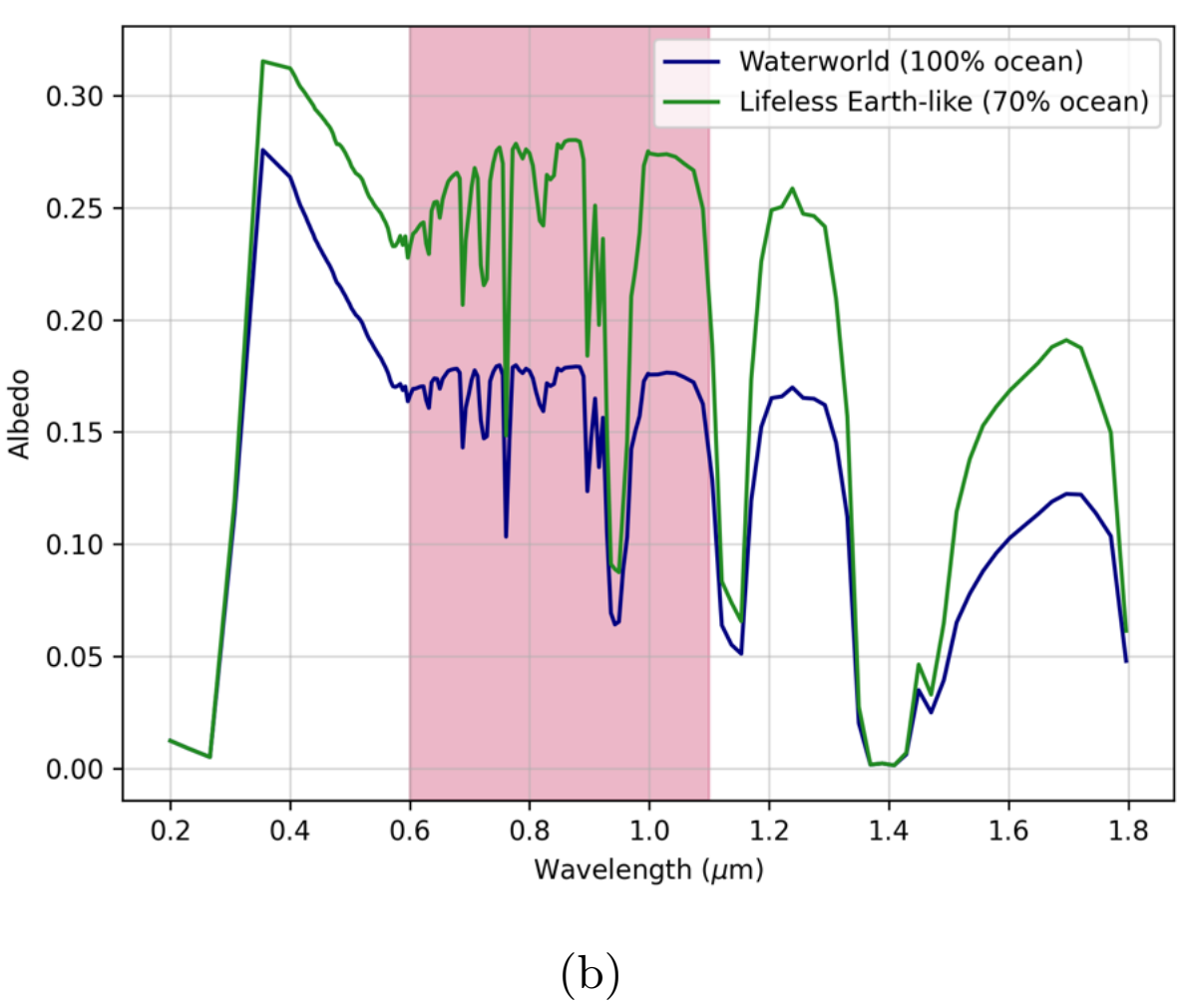

(b)

**Figure 1.** (a) The reflectance spectra of representative lifeless surfaces used in this study. Land surfaces are solid lines, ocean/ice/snow are dashed, and the dotted line is the nominal lifeless and airless Earth spectrum, which is an area weighted sum of the other surfaces as described in the main text. All land surfaces have an upward slope in the UV-to-visible-to-NIR portion of the spectrum, in contrast with the flat ocean spectrum or downward-sloping ice/snow spectra. This difference enables land detection via constraining the shape of the surface spectrum in reflected light. (b) Comparison of a waterworld (blue line) to that of a lifeless Earth-like planet with 30% land (green line), both with an overlying Earth-like atmosphere. Introducing land increases the overall albedo, and creates a slope in the visible-NIR region of the whole-planet spectrum (shaded in pink) that is observationally distinct from a pure ocean surface.

### 2.2. Radiative Transfer Retrieval Model

Our atmosphere and surface retrievals were completed using the open-source 1D radiative transfer and retrieval code `rfast`

**Table 1**
List of Unknown Parameters Used in Retrievals

| Parameter | Symbol | True Value ($\log_{10}$) | Prior Range ($\log_{10}$) |
|---|---|---|---|
| $N_2$ (Pa) | $N_2$ | 4.896 | [−2, 7] |
| $O_2$ (Pa) | $O_2$ | 4.327 | [−2, 7] |
| $H_2O$ (Pa) | $H_2O$ | 2.481 | [−2, 7] |
| $O_3$ (Pa) | $O_3$ | −1.151 | [−2, 7] |
| $CO_2$ (Pa) | $CO_2$ | 1.606 | [−2, 7] |
| CO (Pa) | CO | −1.996 | [−2, 7] |
| $CH_4$ (Pa) | $CH_4$ | −0.695 | [−2, 7] |
| Fractional land coverage | Apars($A_n$) | See Section 2.1 | [−4, 0] |
| Planet radius ($R_\oplus$) | $R_p$ | 0.0 | [−0.5, 0.5] |
| Planet mass ($M_\oplus$) | $M_p$ | 0.0 | [−1, 1] |
| Cloud thickness (Pa) | $\Delta P_c$ | 4.0 | [0, 7] |
| Cloud-top pressure (Pa) | $P_t$ | 4.778 | [0, 7] |
| Cloud optical depth | $\tau_{c0}$ | 1.0 | [−3, 3] |
| Cloud fraction | $f_c$ | −0.301 | [−3, 0] |

**Note.** Rows 1–7 are partial pressures of constituent gases (in $\log_{10}$ Pa). Rows 8–14 represent surface, bulk planet, and cloud property parameters. Columns denote parameter name, symbol, $\log_{10}$ assumed value used to generate the nominal true spectrum, and $\log_{10}$ prior range.

(T. D. Robinson & A. Salvador 2023). Atmospheric species include $N_2$, $O_2$, $H_2O$, $O_3$, $CO_2$, CO, and $CH_4$, and mixing ratios in Earth's modern troposphere are used as true values. We assume constant gas mixing ratios with altitude using Earth-like bulk values, and a 294 K isothermal atmosphere. The associated molecular opacities used in the model are those from the HITRAN2020 database (I. Gordon et al. 2022). Clouds in `rfast` are parameterized by a gray cloud model that includes cloud opacity, cloud-top pressure, cloud thickness, and total cloud fraction. Table 1 shows the parameters to be retrieved, their true values, and our assumed prior ranges. Planetary mass and radius are also unknown parameters in our simulated retrievals. We parameterize the planet surface as a linear combination of plausible rocky and ocean/ice/snow surfaces (see Section 2.1 and Figure 1(a)).

To simulate noisy spectra, `rfast` is equipped with a noise model that applies a fixed noise to each spectral point based on the continuum at 1 $\mu$m. For all retrievals we assumed a resolution, $R = \lambda/\Delta\lambda$, $R = 7$ in the UV ($\lambda = 0.2$–$0.4$ $\mu$m), $R = 140$ in the visible ($\lambda = 0.4$–$1.0$ $\mu$m), and $R = 70$ in the NIR ($\lambda = 1.0$–$1.8$ $\mu$m). Retrievals in `rfast` are handled with Markov Chain Monte Carlo (MCMC) sampling to determine what combination of atmospheric species and surface features is most consistent with the true spectrum (D. Foreman-Mackey et al. 2013). Table 1 outlines our choices of logarithmic uniform priors for all parameters used in our model. Gas abundances are retrieved in partial pressure space, which avoids biased priors compared to retrievals in mixing ratio space. If mixing ratios are retrieved, then a background gas must be assumed with a mixing ratio equal to 1 minus all other constituents. The implied prior for the background gas can then be highly skewed toward larger values (S. Hall et al. 2023). Partial pressure retrievals are generally equivalent to a centered log-ratio approach (B. Benneke & S. Seager 2012; M. Damiano & R. Hu 2021, 2022), except that total pressure is not retrieved as a separate parameter and is instead the sum of all constituent partial pressures.

The number of parameters for the surface albedo model varies with the number of surfaces. In all cases, we run 200 walkers with 200,000 steps, a burn-in of 100,000, and thinning of 10.

### 2.3. Land Fraction Retrievals

Figure 1(b) shows the reflected-light spectra of a planet with an Earth-like atmospheric composition and either zero land (waterworld) or a lifeless Earth-like surface (the linear combination of ocean, ice, and land surfaces described in Section 2.1). The upward-sloping surface spectrum for the lifeless Earth in Figure 1(a) imparts an upward slope in the lifeless Earth whole-planet spectrum of Figure 1(b) starting around 0.6 $\mu$m, when compared to the same atmosphere with an ocean surface (note that the lifeless Earth-like surface also has a greater overall albedo, but this is not necessarily diagnostic of land due to possible degeneracies with cloud properties and planet size). Parameterizing a nongray surface albedo in this way introduces a wavelength-dependent albedo function, $A_s$, where the retrieved parameters ($A_0$, $A_1$, $A_2$, $A_3$, $A_4$) are the land fractions of the respective surfaces (weathered granite, kaolinite, weathered basalt, desert, and ice). In these retrievals the ocean fraction is retrieved implicitly as $1 - (A_0 + A_1 + A_2 + A_3 + A_4)$. This is similar to the surface albedo model of J. Gomez Barrientos et al. (2023), except here we do not prescribe distinct wavelength regions in our parameterization.

### 2.4. Agnostic Retrievals

While terrestrial exoplanets may have broadly similar bulk compositions to the Earth, we cannot rule out more exotic surface compositions. Thus, we attempt to demonstrate land detection without a priori assumptions about rocky surface composition. To do this, we perform retrievals of surfaces parameterized by polynomials which are agnostic to surface composition. We use our nominal lifeless Earth spectrum as truth in the retrieval, and fit either a straight line or a higher-order polynomial. Instead of using `rfast` to retrieve the individual land fractions, $A_n$, as in the previous section, we configure the model to instead retrieve on the coefficients of the polynomial. In this way, we demonstrate that land can be uniquely identified by its spectral behavior without knowing its exact composition prior to observation, i.e., by showing that the retrieved surface spectrum is distinct from any combination of water/ice/snow. Table 2 outlines the structure of how the agnostic retrievals are parameterized, including our imposed prior ranges.

### 2.5. Coronagraph Noise Model

We estimate the observation times required to reach a certain SNR for a land detection using the reflected-light coronagraph noise model `coronagraph` (J. Lustig-Yaeger et al. 2019). This is accomplished by splitting up the `rfast` spectra into resolution bins according to the resolution, $R$, from Section 2.2, and calculating associated observation times. This coronagraph model includes the following sources of noise: read, speckles, dark current, thermal, zodiacal, exozodiacal, and clock-induced charge. We repeated this process for each spectrum with two different telescope diameters: 6 and 8 m, based on possible architectures for HWO (National Academies of Sciences 2023). For each observing-time calculation, we assumed the exoplanet is Earth-like in radius and mass, and is 10 pc away (E. Mamajek & K. Stapelfeldt 2023).

**Table 2**
The Retrieved Coefficients and Prior Ranges for Each of Our Agnostic Surface Spectra Parameterizations: Linear, Parabolic, and Cubic

| Function | Coefficients | Prior Ranges |
|---|---|---|
| Linear | $A_0$, $A_1$ | $A_0$: $[-\frac{1}{1.6}, \frac{1}{1.6}]$ |
| | … | $A_1$: [−5, 1] |
| | … | $A_0$: [−5, 5] |
| Parabolic | $A_0$, $A_1$, $A_2$ | $A_1$: [−5, 5] |
| | … | $A_2$: [−5, 5] |
| | … | $A_0$: [−5, 5] |
| | … | $A_1$: [−5, 5] |
| Cubic | $A_0$, $A_1$, $A_2$, $A_3$ | $A_2$: [−5, 5] |
| | … | $A_3$: [−5, 5] |

**Note.** Note that the parameters for each function are retrieved in linear space.

## 3. Results

### 3.1. Lifeless Earth-like Planets

#### 3.1.1. Directly Retrieving Surface Land Fractions

The retrieved spectrum from our simulated nominal lifeless Earth retrieval is shown alongside the true input spectrum in the inset of Figure 2. The 95% confidence interval for the retrieved spectrum falls within the uncertainty of the true spectrum, suggesting convergence (this is also supported by inspection of walkers). All retrieved parameters, including individual surface fractions (i.e., weathered basalt, granite, kaolinite, ice, desert), are shown in Figure 2. Consistent with previous simulated retrieval studies of Earth analogs (Y. K. Feng et al. 2018; M. Damiano & R. Hu 2021, 2022; E. Alei et al. 2022; J. Gomez Barrientos et al. 2023; N. Latouf et al. 2023, 2024; S. Gilbert-Janizek et al. 2024), we find that atmospheric $O_2$, $H_2O$, $O_3$, and total pressure are all well constrained by the reflected-light spectrum despite the unknown surface composition and Earth-like cloud coverage. Planetary mass and radius are poorly constrained, and only upper limits on atmospheric CO, $CO_2$, and $CH_4$ are possible for Earth-like abundances. We show the total land fraction posterior—that is, weathered basalt+weathered granite+kaolinite+desert—for this nominal SNR 20 simulation as well as SNR 10 and 40 cases in Figure 3(a). The posteriors indicate that a confident land detection requires a minimum SNR of 20.

The nominal retrieval described above assumes an optimistic wavelength range of 0.2–1.8 $\mu$m. To test the robustness of this result against restricting the wavelength range, we repeated the nominal simulation at an SNR of 20 for a selection of more restrictive wavelength ranges, shown in Figure 3(c). Our most restrictive simulation of 0.4–0.7 $\mu$m results in a nondetection of land. Extending in both long and short wavelengths to 0.3–1.1 $\mu$m is the narrowest wavelength range that results in a robust land detection for an SNR of 20. Figure 3(c) additionally demonstrates the importance of including the UV down to a minimum of 0.3 $\mu$m in order to rule out a waterworld confidently at an SNR of 20. This is because more UV spectral information breaks a degeneracy between high $O_3$ and low land due to their similar spectral slopes—Chappuis band $O_3$ absorption around 0.4–0.65 $\mu$m imparts a U-shape to Earth's reflectance spectrum (J. Krissansen-Totton et al. 2016) that can mimic the upward slope of land in the visible, and so UV coverage is needed to independently constrain $O_3$. This is demonstrated in Figure 4, where 0.4–1.1 $\mu$m results in a low land–high $O_3$ degeneracy. We find that at an SNR of 20, the posteriors for total land fraction are unaffected by the long-wavelength cutoff, provided it extends to at least 1.1 $\mu$m, as shown in Figure 3(d).

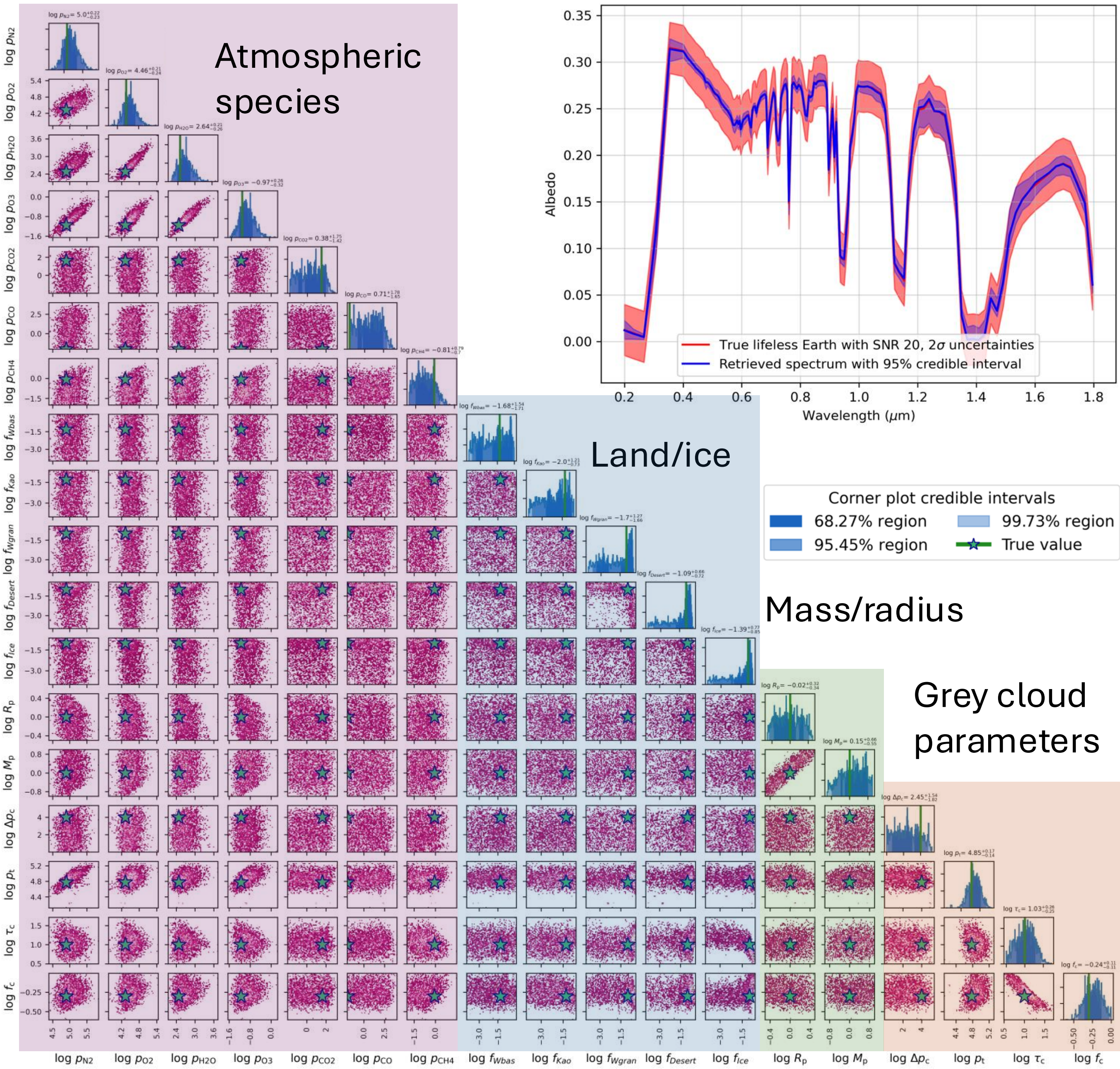

**Figure 2.** Posteriors for atmospheric and surface retrieval for our nominal lifeless Earth using `rfast`. Corner plot shows the 2D marginalized posterior distributions for all pairs of parameters. Diagonal distributions show the 1D marginalized posteriors for each parameter with 68.27%, 95.45%, and 99.73% credible regions denoted by shades of blue (P. Cubillos et al. 2017). Above each diagonal distribution we list the retrieved value with 68.27% credible intervals. The green vertical lines and green stars denote the true values. Table 1 summarizes the model parameters of this corner plot in the order they are plotted here. Inset: reflectance spectrum of true lifeless Earth with 30% land with assumed SNR 20 uncertainty ($R = 7$ in UV, $R = 140$ in visible, $R = 70$ in NIR) alongside the retrieved spectrum in blue with shaded 95% credible interval.

The sensitivity of the land detection to wavelength range is highly reduced at an SNR of 40. We find that land can still be constrained even at the most restrictive wavelength range of 0.4–0.7 $\mu$m, and including the UV to at least 0.3 $\mu$m is no longer a requirement, as shown in Figure 5(b). Note, however, that it is unlikely SNR 40 would be achievable in a feasible observing time for all HWO targets, and so the ability to detect land at lower assumed SNRs remains an important consideration for nominal wavelength coverage.

We also test the sensitivity of our retrievals against increasing cloud fraction and decreasing total land fraction. Figure 3(b) plots the posteriors for different assumed cloud fractions. Unsurprisingly, a cloud fraction of 100% results in a nondetection. At an SNR of 20, a confident land detection is possible for cloud fractions of approximately 50%–60% or less, where 50% clouds is our nominal scenario, also shown in Figure 3(a). We performed the same sensitivity test for a higher SNR of 40 and found that for every cloud fraction the total land posteriors peak at a higher probability density compared to an SNR of 20, indicating a more confident detection (Figure 5(a)). Similarly, in Figure A1, we keep the cloud fraction constant at 50% for an SNR of 20 and decrease the total land fraction. Based on these simulations, a land detection given our imposed cloud and SNR requires a land fraction greater than 25%. With 50% cloud coverage, we determined the minimum land fraction required for detection at an SNR of 40 is greater than 15% (Figure A2).

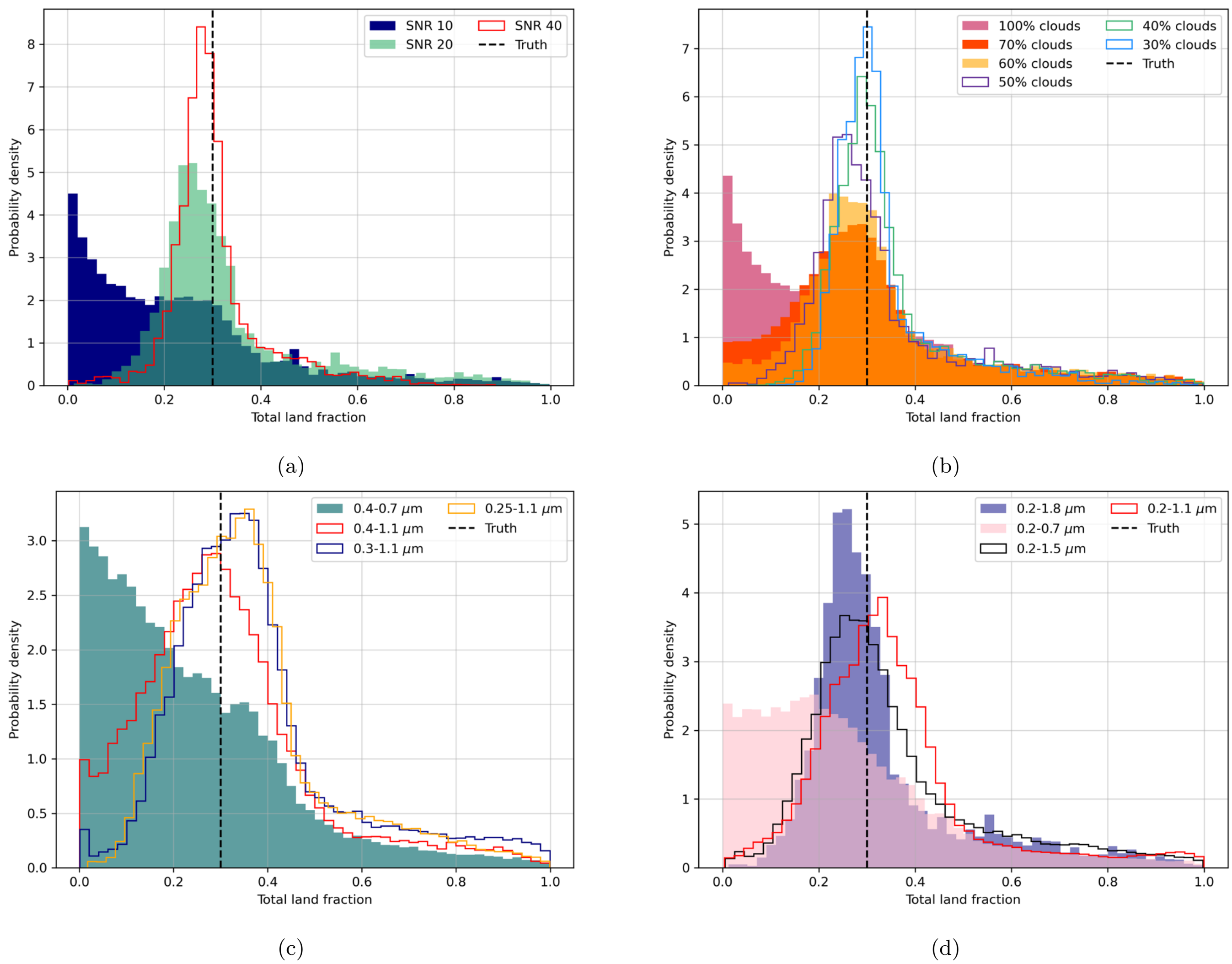

**Figure 3.** (a) Total land fraction posteriors for the nominal Earth-like retrieval at different SNRs with 50% cloud coverage. Given these conditions, an SNR of 20 is required for land detection. (b) Total land fraction posteriors for the nominal Earth-like retrieval at an SNR of 20 with varying cloud fractions from 30% to 100%. One-hundred percent results in a clear nondetection, and for an SNR of 20, cloud fractions of approximately 50%–60% or less are necessary for a land detection. (c) Total land fraction posteriors for the nominal Earth-like retrieval at an SNR of 20 for more restrictive wavelength ranges. A wavelength coverage of at least 0.3–1.1 $\mu$m is required for detection (dark blue contour). Restricting to 0.4–1.1 $\mu$m (red contour) results in a nondetection. (d) Long-wavelength cutoff tests for an SNR of 20. We find the total land constraint to be relatively unaffected by the cutoff as long as it is beyond 1.1 $\mu$m (red contour). The pink contour, 0.2–0.7 $\mu$m, does not permit land detection.

### *3.1.2. Agnostic Retrievals*

Retrievals in the previous section were based on a priori assumptions about what surfaces make up the land on a planet (e.g., basalt, granite, etc.). In reality, we should expect to have no knowledge of surface composition, and must instead detect land with a more agnostic approach. To do this, we fit the nominal lifeless Earth simulation with polynomials and retrieve on the coefficients instead of the land fractions (see Table 2). Figure 6(a) shows our nominal "true" lifeless Earth spectrum in purple as well as the retrieved spectra for each of the three surface parameterizations: linear, parabolic, and cubic.

Figure 6(b) overplots the retrieved surface spectra for our linear, parabolic, and cubic parameterizations, in addition to the true lifeless Earth surface with 30% land. The linear fit captures the basic principle behind this method: A surface containing land is statistically distinct from a 100% ocean surface because land surfaces have positive slopes in the visible-to-NIR. All three of our tested parameterizations are inconsistent with a surface composed of 100% ocean, which demonstrates the retrieval requires a surface resembling nonzero total land regardless of what that land is composed of, and independent of the surface parameterization used to fit the data.

In order to completely exclude a waterworld false positive, a 100% snow/ice/ocean mixture must also be ruled out. As such, Figure B1 plots the same three surface parameterizations as well as surfaces composed of linear combinations of ice and water in the dashed lines. None of these ice-water mixtures are consistent with the agnostically retrieved surfaces, demonstrating that land is needed to explain the observed spectrum.

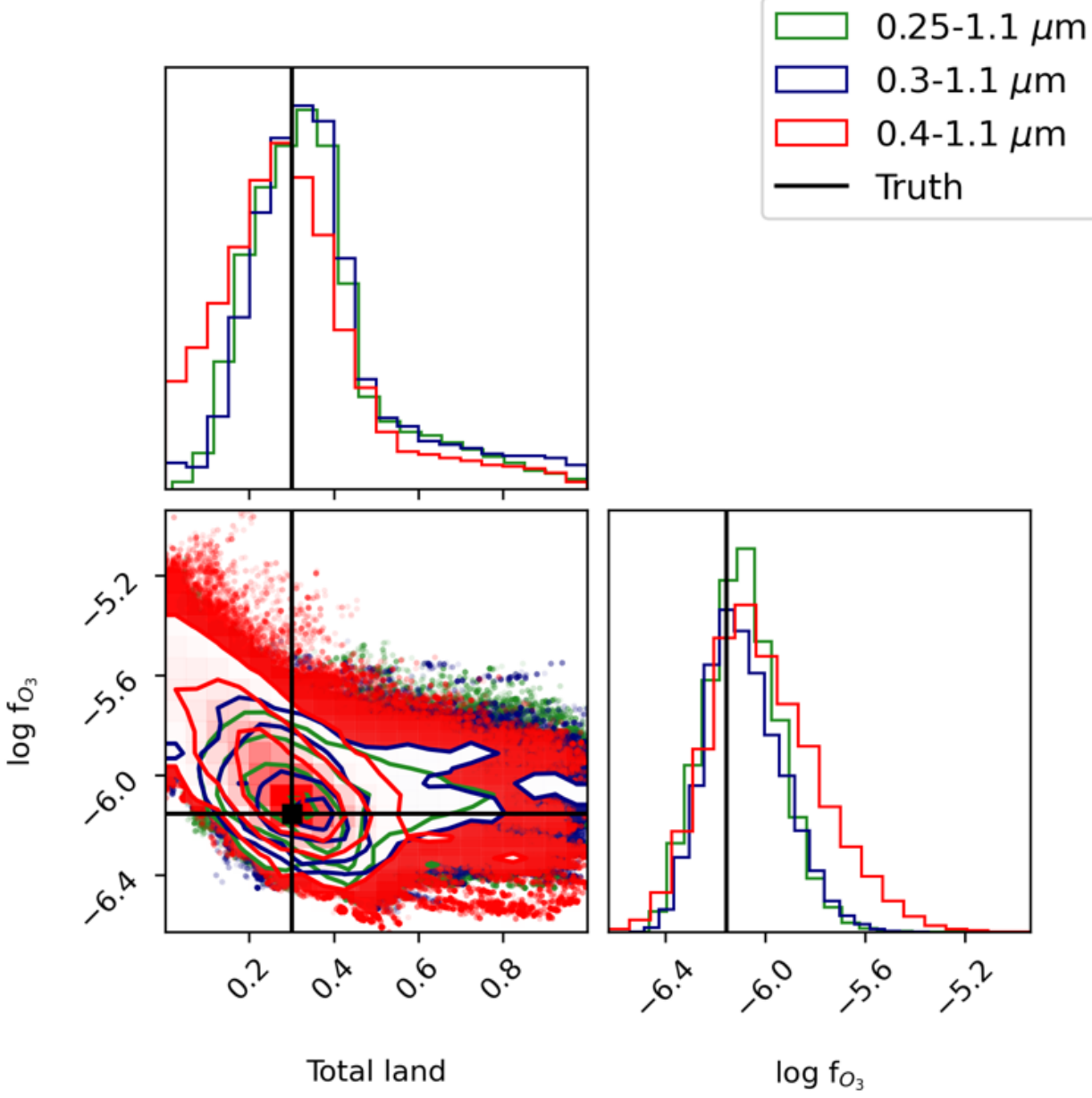

**Figure 4.** Marginalized posteriors for total land fraction and $O_3$ for three lifeless exo-Earth retrievals with different short-wavelength cutoffs. The posteriors for a 0.4–1.1 $\mu$m cutoff (orange) demonstrate the low land–high $O_3$ degeneracy that precludes a confident land detection. The blue contour shows the minimum wavelength requirement for land detection: 0.3–1.1 $\mu$m.

### 3.2. Observing-time Estimates

Our surface land fraction retrievals demonstrate an Earth-like land fraction can be constrained with a minimum SNR of 20, and a wavelength coverage of 0.3–1.1 $\mu$m. Using the telescope noise simulator `coronagraph`, we can calculate how many integration hours HWO would require to make this observation, optimistically assuming all wavelengths can be observed simultaneously. For a 6 m telescope, an SNR of 20 across the visible can be reached in 538 hr. Increasing the aperture to 8 m decreases this time to 211 hr, as shown in Figure 7. Reaching an SNR of 40 requires 2100 and 844 hr for a 6 and 8 m aperture, respectively (not shown).

## 4. Discussion

### 4.1. Implications for Telescope Design

The land detection method demonstrated here provides an avenue for confidently ruling out waterworld false positives given sufficient SNR and wavelength coverage. Our retrievals suggest that, for Earth-like overlying atmospheres and clouds, emerged land can be confidently detected with a minimum wavelength coverage of 0.3–1.1 $\mu$m at an SNR of 20 (assuming $R = 7$ in UV, $R = 140$ in visible, and $R = 70$ in NIR). Crucially, including the UV at least to 0.3 $\mu$m is critical for achieving a land detection since this breaks a degeneracy in the retrieval between the apparent forcing of a red slope by blueward absorption from the ozone Chappuis band and a true upward slope from land properties.

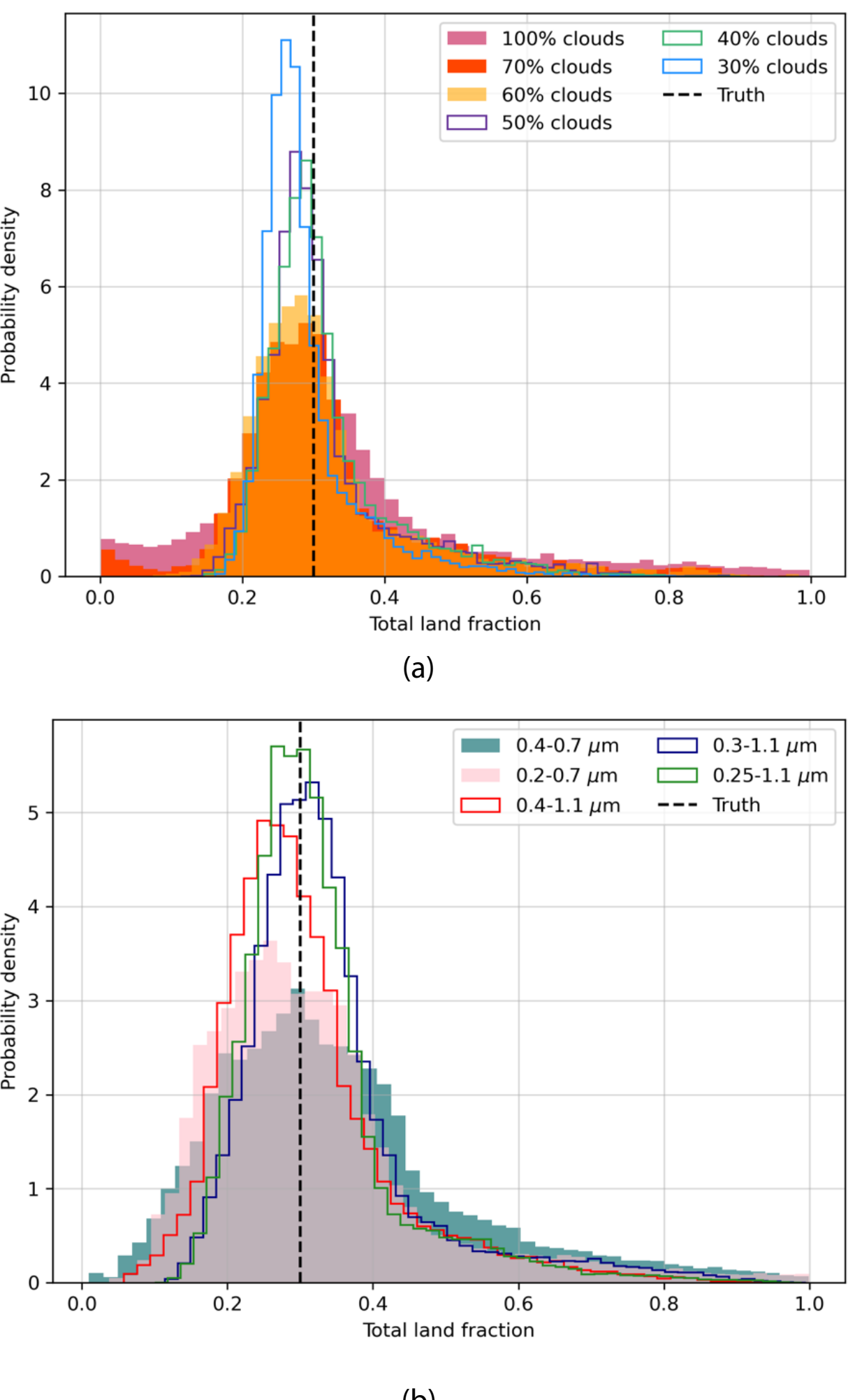

**Figure 5.** (a) Same as Figure 3(b) except with a higher SNR of 40. Here, the land fraction posteriors peak at higher probability density for every cloud fraction compared to the SNR 20 case, indicating a more confident detection. (b) Same as Figure 3(c) except with SNR 40. Land fraction posteriors for a sample of wavelength restrictions for a higher SNR. We find the land fraction constraint to be unaffected by the wavelength range.

To obtain the requisite SNR of 20, assuming an optimistic 25% bandpass for HWO, spectral characterization across the minimum wavelength coverage of 0.3–1.1 $\mu$m with a 6 m telescope would take ∼2100 hr (nearly 3 months) to reach an SNR of 20. A 3 month integration time is likely unrealistic given possible stability issues, overheads, and challenges associated with continuum offsets between observations as the phase angle is changing. With an 8 m aperture the same observation could be made in only ∼840 hr, which is comparable to the Hubble Extreme Deep Field (∼540 hr). For the prescribed 25 exoplanets in HWO's survey this would take ∼2.5 yr. For a 6 m aperture a full survey would take 6–7 yr. These results highlight the need for a larger aperture for HWO to detect land and rule out waterworld false positives in a more reasonable timeframe.

### 4.2. Comparison with Previous Surface Characterization Studies

Similar to J. Gomez Barrientos et al. (2023), we have demonstrated that surface features can be inferred in a retrieved spectrum if a nongray surface albedo model is adopted alongside the atmosphere. Their parametric surface model is used to retrieve the spectral feature of the sharp vegetative red edge, whereas we independently constrain the fractions of each land surface, which has the advantage of being able to identify the less pronounced changes in albedo associated with ruling out a waterworld false positive. This method does require prior assumptions about possible surface composition. In contrast, our agnostic approach makes no a priori assumptions about surface composition, and thus enables confident land detection by contrast with all possible water surfaces.

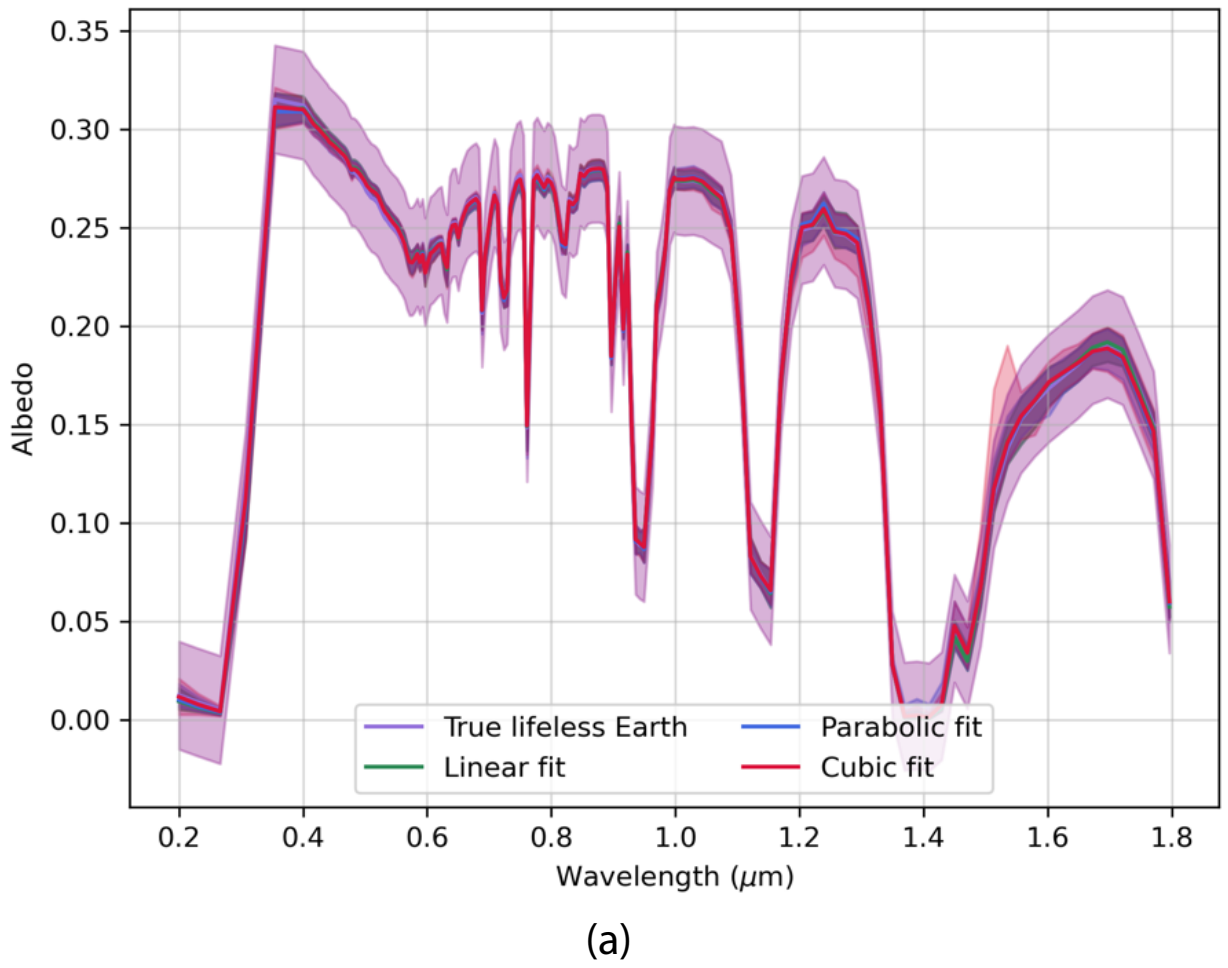

(a)

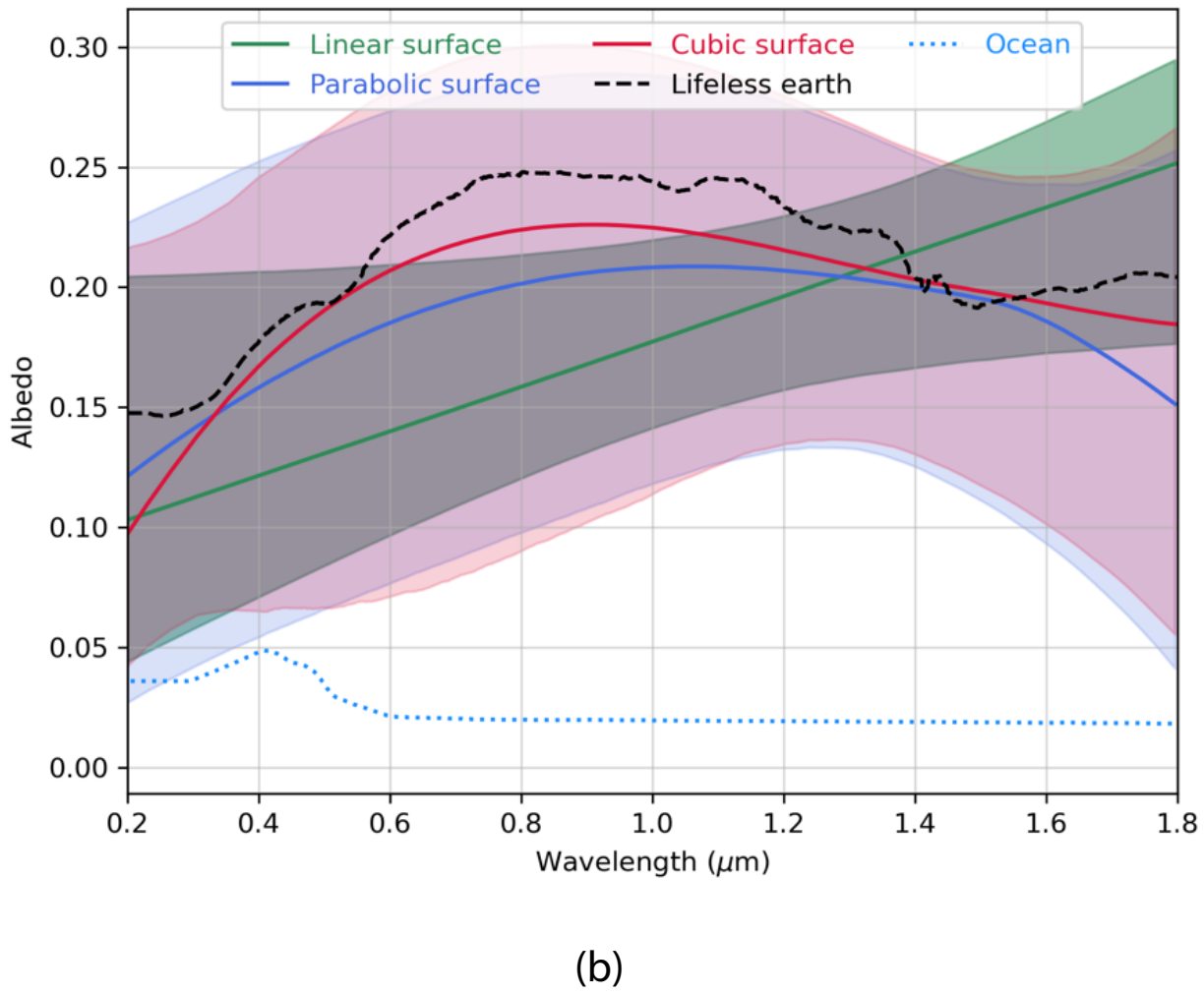

(b)

**Figure 6.** Retrieved spectra and surface spectra for agnostic retrieval that makes no assumptions about crustal composition. (a) The same nominal lifeless whole-Earth spectrum (purple line with SNR 20 uncertainty) is shown alongside the retrieved spectra for the linear, parabolic, and cubic agnostic fits in green, blue, and red, respectively. Excluding a slight overshoot by the cubic fit at 1.5 $\mu$m, all three confidence intervals fall within the uncertainties of the true lifeless Earth shown by the purple envelope. (b) Retrieved surface spectra for our polynomial parameterizations. The true lifeless Earth surface is shown by the black dashed line, and that of pure ocean shown by dotted blue. None of the 95% confidence intervals of the retrieved surfaces are compatible with pure ocean, indicating a waterworld has been ruled out.

The technique described in this study would be applicable to all single-epoch observations with HWO, including the more brightly illuminated gibbous phases, since it does not have any phase dependence. Spatial mapping (N. B. Cowan et al. 2009; H. Kawahara & Y. Fujii 2010, 2011), ocean glint (specular reflection; T. D. Robinson et al. 2010; D. J. Ryan & T. D. Robinson 2022), and spatial mapping of ocean glint

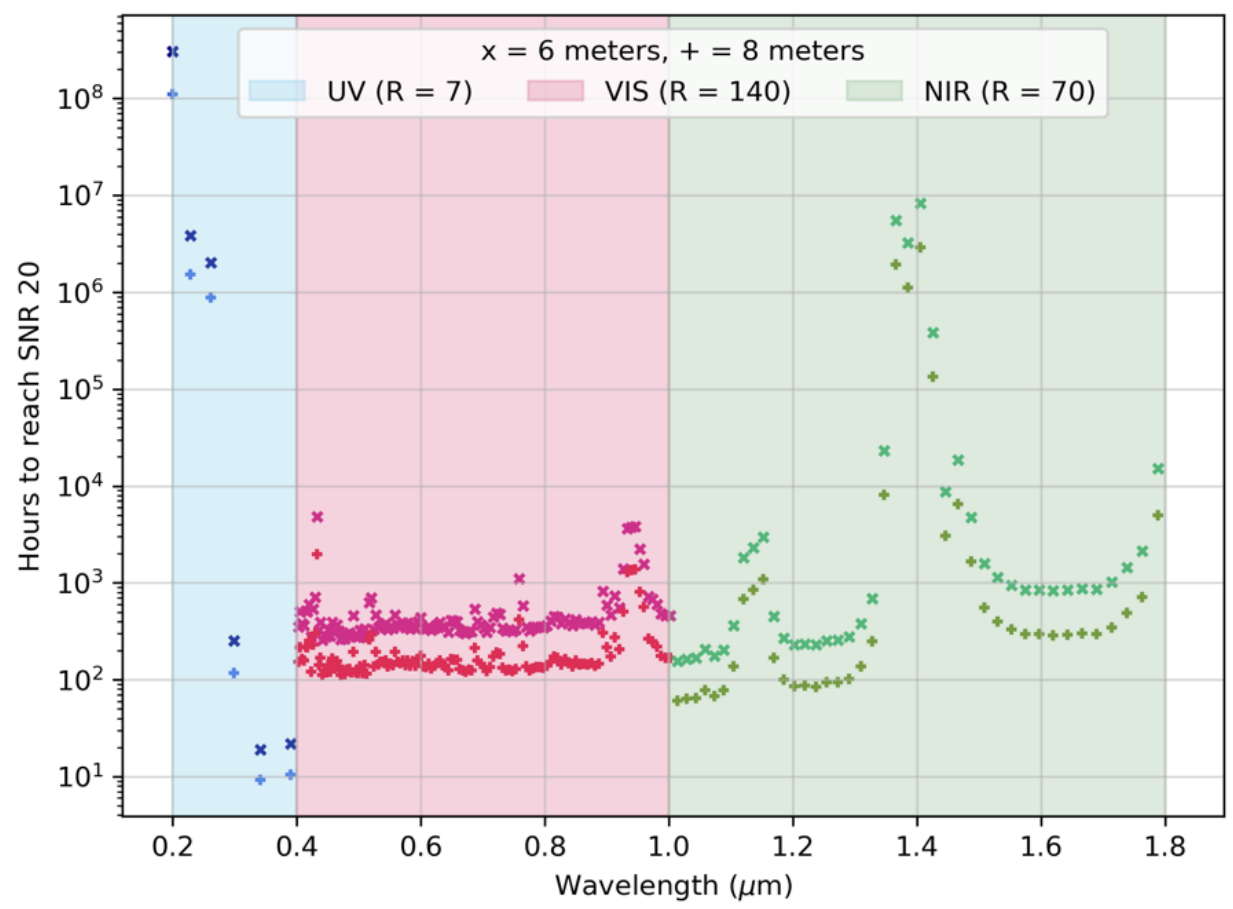

**Figure 7.** Observing-time calculations to reach an SNR of 20 for our nominal lifeless Earth spectrum. The wavelength regions are split according to resolution and calculations for 6 and 8 m apertures are shown by the $\times$ and $+$ symbols, respectively. In the region of interest, 0.3–1.1 $\mu$m, an SNR of 20 can be reached in 538 hr with a 6 m aperture, and 211 with an 8 m aperture.

(J. Lustig-Yaeger et al. 2018) all require observing-time-dependent heterogeneities of the surface potentially attributable to oceans or continents. Additionally, mapping ocean glint requires observations at crescent phase which may not be accessible to a large fraction of HWO targets due to inner-working-angle constraints. Furthermore, observations at crescent phase, where the planetary flux is lowest, would require much longer integration times than those calculated for this method. Consequently, the method presented here for observing continents in a single observation with HWO could be leveraged as an initial characterization tool to identify worthwhile targets to follow-up on with more targeted observations.

Our own investigation of ocean detection is shown in Figure C1. For 0.3–1.1 $\mu$m (and longer), the posterior is suggestive of the presence of an ocean even for an SNR of 20. At a higher SNR of 40, ocean is suggested at even the restricted wavelength range of 0.4–0.7 $\mu$m, as shown in Figure C2. For both SNR cases, the long tails extending toward zero ocean fraction means there is potential ambiguity in the detections, but likewise identifies targets to follow-up on with time-resolved observations.

### 4.3. Sensitivity to Surface Composition

Our nominal retrieval for an Earth-like planet assumed the surface basalt and granite is physically weathered, resulting in finer-grained rocks. Using spectra for solid, unweathered granite and basalt in our retrievals decreases the positive slope of the visible spectrum, and decreases the significance of the land detection (Figure A3). However, fine-grained, weathered spectra are more realistic since Earth-like exoplanets with oceans and exposed land will experience an active hydrological cycle and aeolian processes; signs of weathering and erosion are widespread on Mars even though it likely only possessed a transiently active hydrological cycle (A. Gaudin et al. 2011). Our nominal retrieval also includes ice as opposed to snow, which has, on average, higher albedo (see Figure 1(a)). In a retrieval done with snow instead of ice, we find no significant change to the resulting land detection (not shown).

### 4.4. Practical Observing Considerations

The whole-disk average spectrum that HWO sees will be an integration of the land fraction on the visible hemisphere, weighted by the phase angle and the planetary obliquity (N. B. Cowan et al. 2011). In our 1D simulations, we are assuming that this is equivalent to the planet's total land fraction, but this may not be true depending on viewing geometry and rotation rate. For example, viewing Earth from above biases the total land fraction higher than 30% because the Northern Hemisphere has more continents than the Southern. In the worst-case scenario, if all the planetary land is on the hidden hemisphere, then HWO could observe an apparent false positive that was actually inhabited, and corroborating life would likely require detecting coexisting reducing gases with more sensitive future instruments.

Our retrievals also assume that a habitable exo-Earth exists in isolation. Although HWO will spatially resolve exoplanets from their host star, a planet and its moons will typically remain unresolved. Based on solar system observations, we know moons can rival their host planets in certain spectral bands (e.g., Earth's Moon at thermal wavelengths; T. D. Robinson 2011). As such, a blended spectrum with significant contributions from an exomoon is likely to differ from an exoplanet in isolation and might even problematically mimic a nonzero land fraction. This highlights the need for methods to detect and differentiate exomoons around Earth analogs with HWO (E. Agol et al. 2015; M. A. Limbach et al. 2024).

Our retrievals assumed an Earth-like land fraction of 30% and cloud fraction of 50%. At an SNR of 20, we find that land detection is limited to cases with over 25% land, and up to approximately 50%–60% clouds (at an SNR of 40, the land fraction threshold is greater than 15% land, and the peaks of the posteriors increase for every cloud fraction). Above this cloud threshold, land detection would be challenging. On modern Earth, clouds with optical depth >2 cover 56% of the total surface area on average, but there is 10%–15% more cloud cover over the ocean (C. J. Stubenrauch et al. 2013), and so land on an Earth-like exoplanet would be comparatively detectable. Future work ought to explore a wider range of physically motivated land and cloud fractions.

## 5. Conclusion

Detecting land via reflected-light spectroscopy can help HWO rule out $O_2$ biosignature false positives associated with the suppression of $O_2$ sinks due to extremely deep oceans. Land detection is possible because all likely land surfaces for exo-Earth analogs have a positive sloping reflectance spectrum in the visible, whereas liquid water and water ice/snow are flat or slope negatively, respectively. Our reflected-light retrievals presented here suggest that detecting an Earth-like land fraction of 30% would be possible provided HWO has a wavelength coverage of 0.3–1.1 $\mu$m, and can achieve an SNR of 20 on the planetary spectrum (with $R = 140$ in the visible, and $R = 7$ in the UV). Including the UV to at least 0.3 $\mu$m is critical to break the low land–high $O_3$ degeneracy. The minimum SNR of 20 is achievable with 538 hr using a 6 m telescope aperture. This observing time decreases to 211 hr using an 8 m. These SNR and resolution requirements imply a larger aperture will enable land detection for a larger fraction of HWO targets with shorter integration times.

## Acknowledgments

This work was supported by NASA Astrophysics Decadal Survey Precursor Science grant No. 80NSSC23K1471. This work was additionally supported by the Virtual Planetary Laboratory, a member of the NASA Nexus for Exoplanet System Science (NExSS), funded via the NASA Astrobiology Program grant No. 80NSSC23K1398. We would like to thank Jacob Lustig-Yaeger, Amber Young, and the anonymous reviewer for helpful comments.

## Data Availability

The modified version of `rfast` used in this paper can be found on the lead author's Github (https://github.com/AnnaUlses/rfast_surface), and is archived in Zenodo (A. G. Ulses et al. 2025). Similarly, we have archived the data outputs from this paper in Zenodo at doi: 10.5281/zenodo.15678385. These data can be used to reproduce the figures from this publication.

## Appendix A
## Land, Cloud, and Surface Composition Sensitivity Tests

We varied the land fraction and cloud fraction for both SNR 20 and 40 cases and repeated our lifeless Earth nominal retrievals. Figure A1 shows the total land fraction posterior for our nominal Earth-like retrieval at an SNR of 20 and cloud fraction of 50%. Under these conditions, land detection requires a fraction greater than 25%. Figure A2 is similar to Figure A1 except the SNR is increased to 40. Here, a land detection requires greater than 15% fractional coverage. Similarly, we varied the cloud fraction at an SNR of 40. Our nominal simulations included both weathered basalt and granite as opposed to solid, which increases the prominence of the surface land slope in the visible-NIR. Figure A3 demonstrates that using spectra for weathered rocks results in a more constrained total land fraction compared to unweathered.

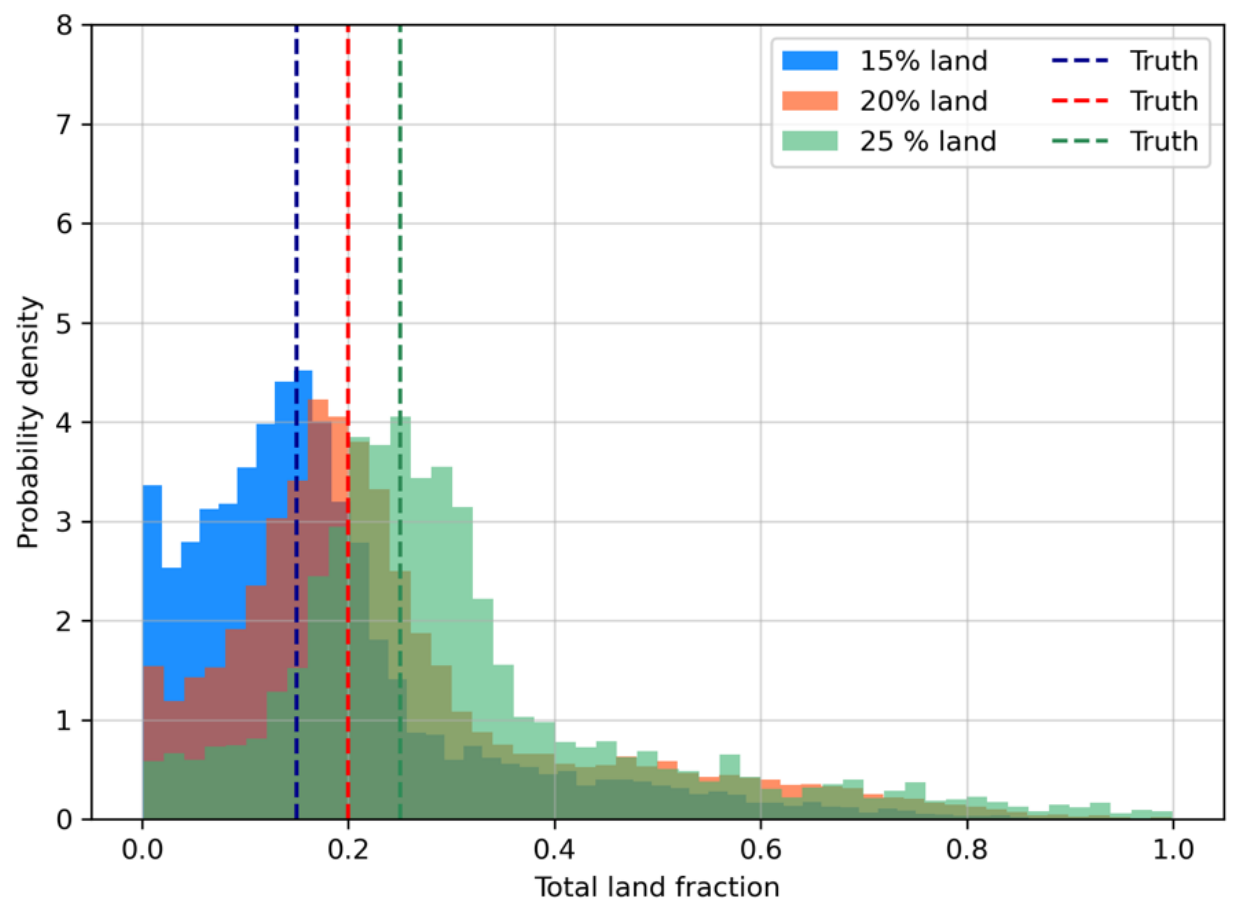

**Figure A1.** Total land fraction posteriors for an Earth-analog retrieval at an SNR of 20 for decreasing total land. With an SNR of 20 and a cloud fraction of 50%, land detection requires a fraction greater than approximately 25%.

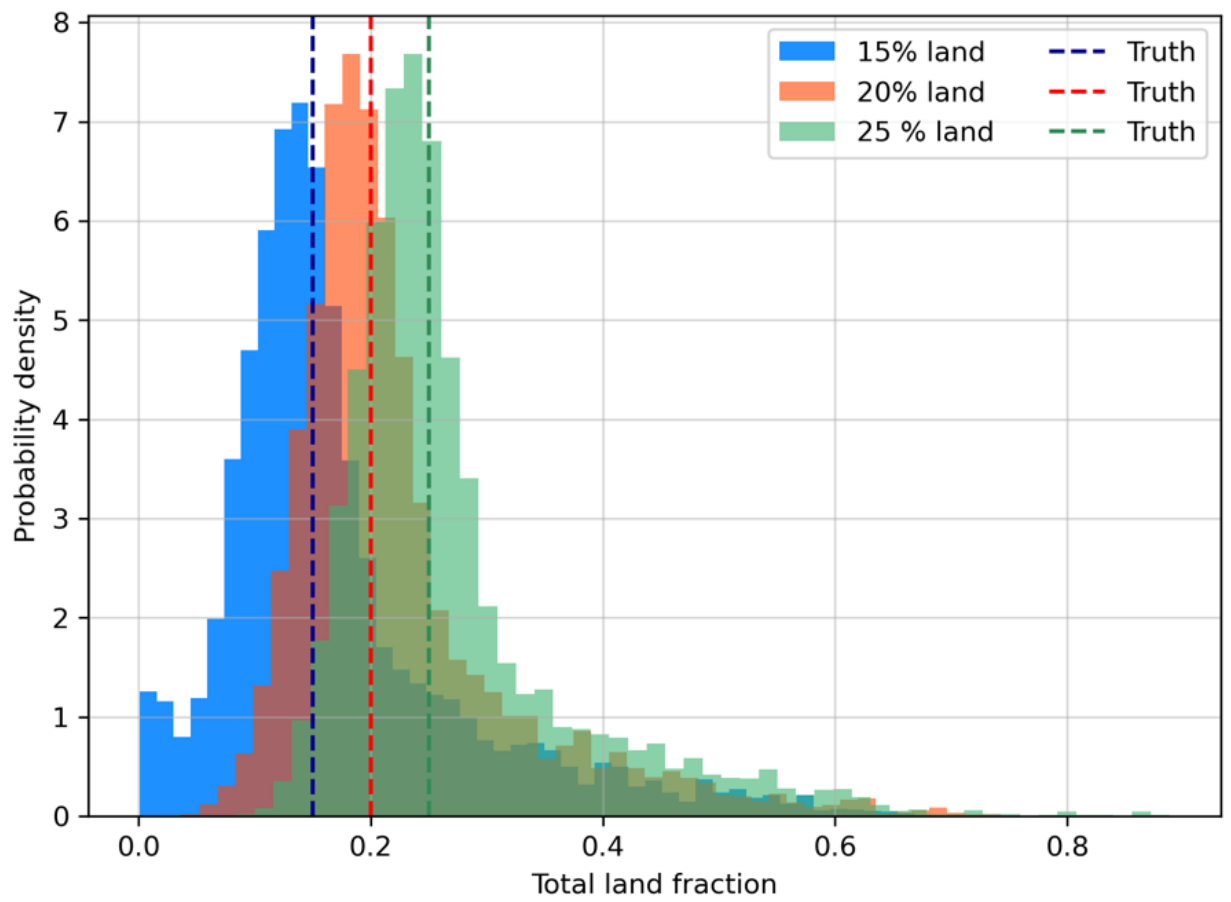

**Figure A2.** Same as Figure A1, except with a higher SNR of 40. The minimum total land coverage for detection is lowered from 25% to greater than 15%.

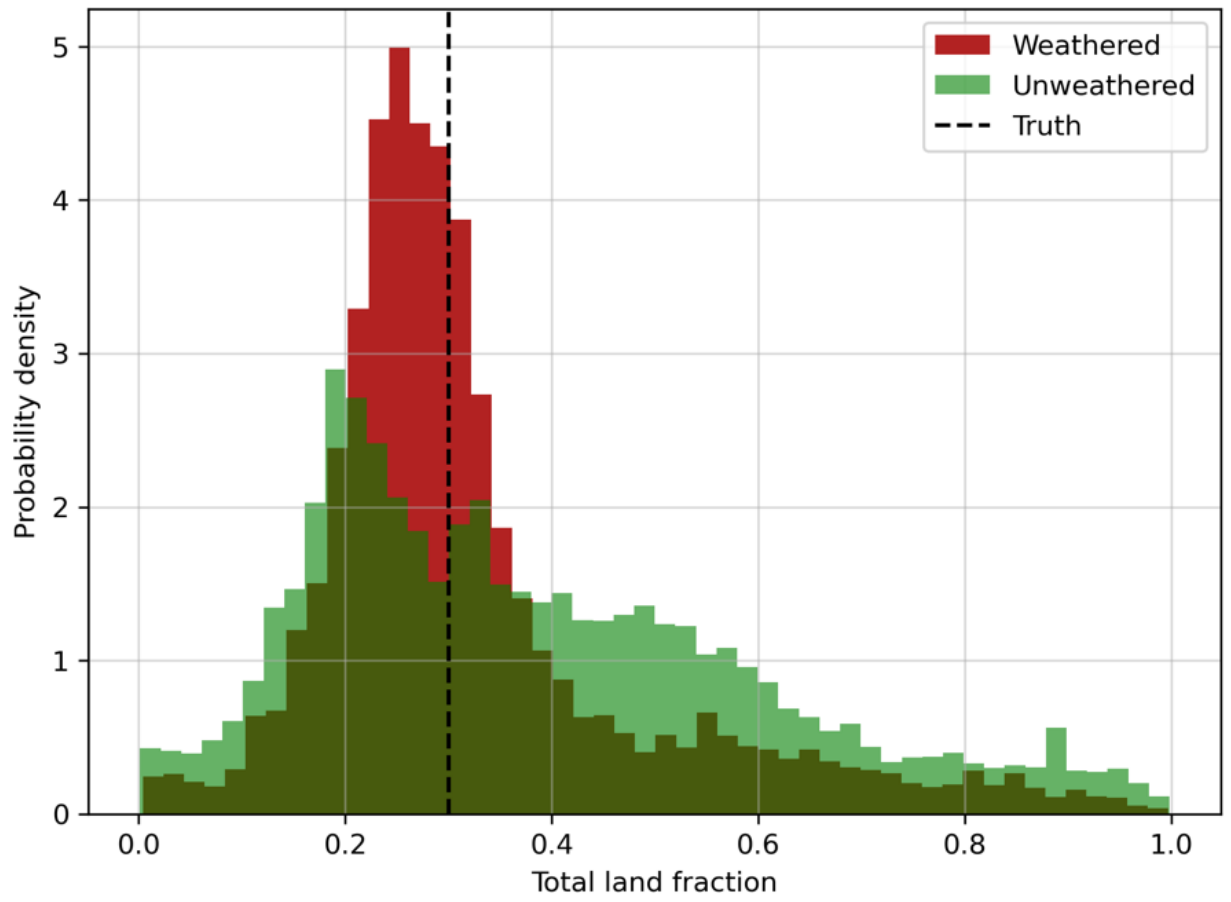

**Figure A3.** Comparison of the total land posterior when using weathered basalt and granite (red distribution) vs. unweathered, solid basalt and granite (green). Weathered rocks have higher albedo and more pronounced slope in the visible-NIR, and so result in a more constrained land fraction.

## Appendix B
## Agnostic Retrievals

As discussed in the main text, our agnostic retrieval provided surface spectra that clearly rule out 100% liquid-water surfaces. Figure B1 shows the same retrieved surface spectra as in Figure 6(b) alongside ice-ocean linear mixtures. Across the full spectral range, our retrieved surfaces are additionally clearly inconsistent with surfaces composed purely of ice and ocean.

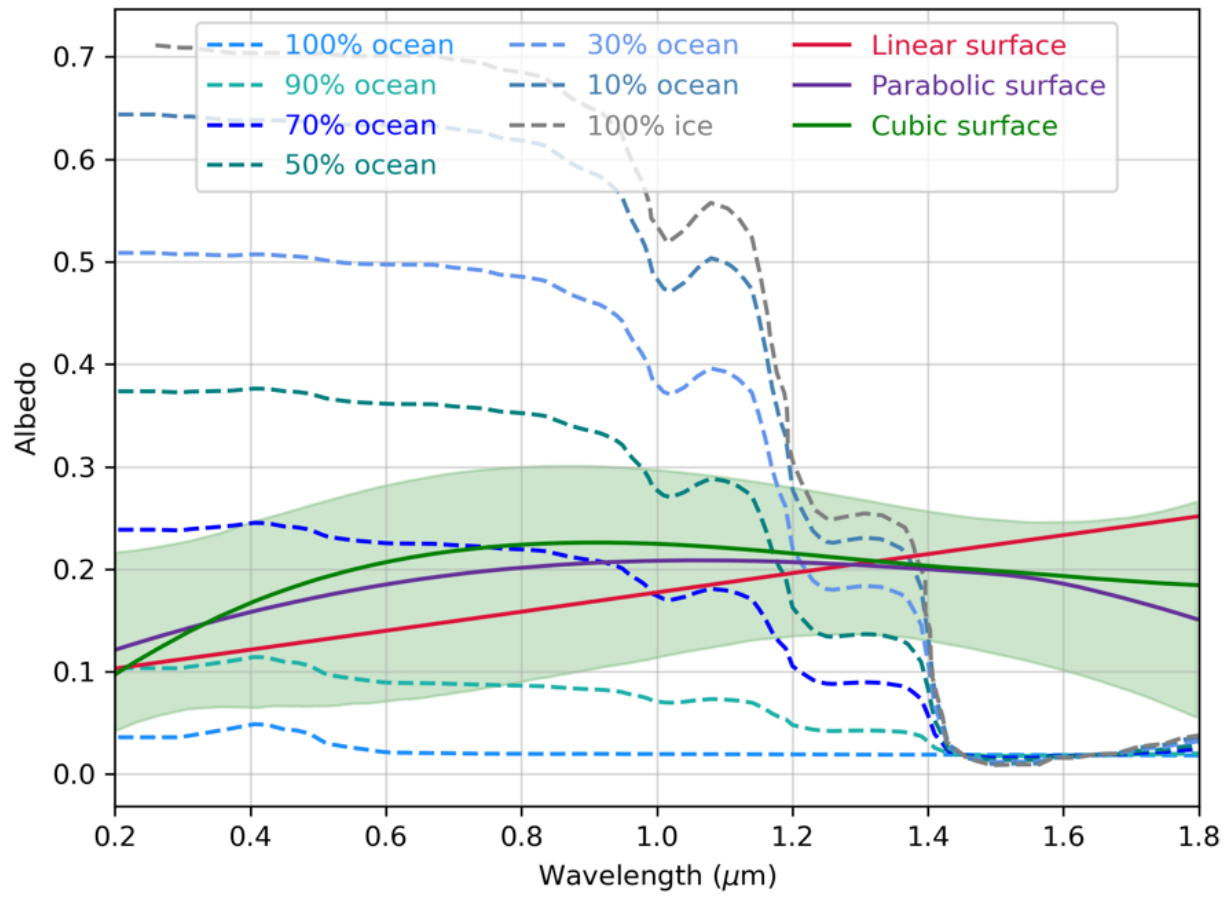

**Figure B1.** The same surfaces shown in Figure 6(b) overplotted with spectra of linear mixtures of ice and ocean surfaces for comparison. Ice/ocean surfaces are clearly incompatible with the retrieved surfaces, suggesting that a waterworld can be ruled out even if the phase of the surface water is unknown.

## Appendix C
## Ocean Detections

In our direct surface fraction retrievals described in Section 2.3, the ocean fraction is retrieved implicitly as 1 – (the sum of all land fractions and ice). Figures C1 and C2 plot the ocean posteriors for SNR 20 and 40, respectively. At an SNR of 20, the ocean fraction is better constrained with a wavelength coverage of 0.3–1.1 $\mu$m and wider; 0.4–0.7 $\mu$m is insufficient. At an SNR of 40, the detection is less dependent on wavelength coverage and is better constrained. Note, as well, that the long tails of the posteriors toward zero ocean are less pronounced at this higher SNR, which suggests ocean could be detected with longer exposure times.

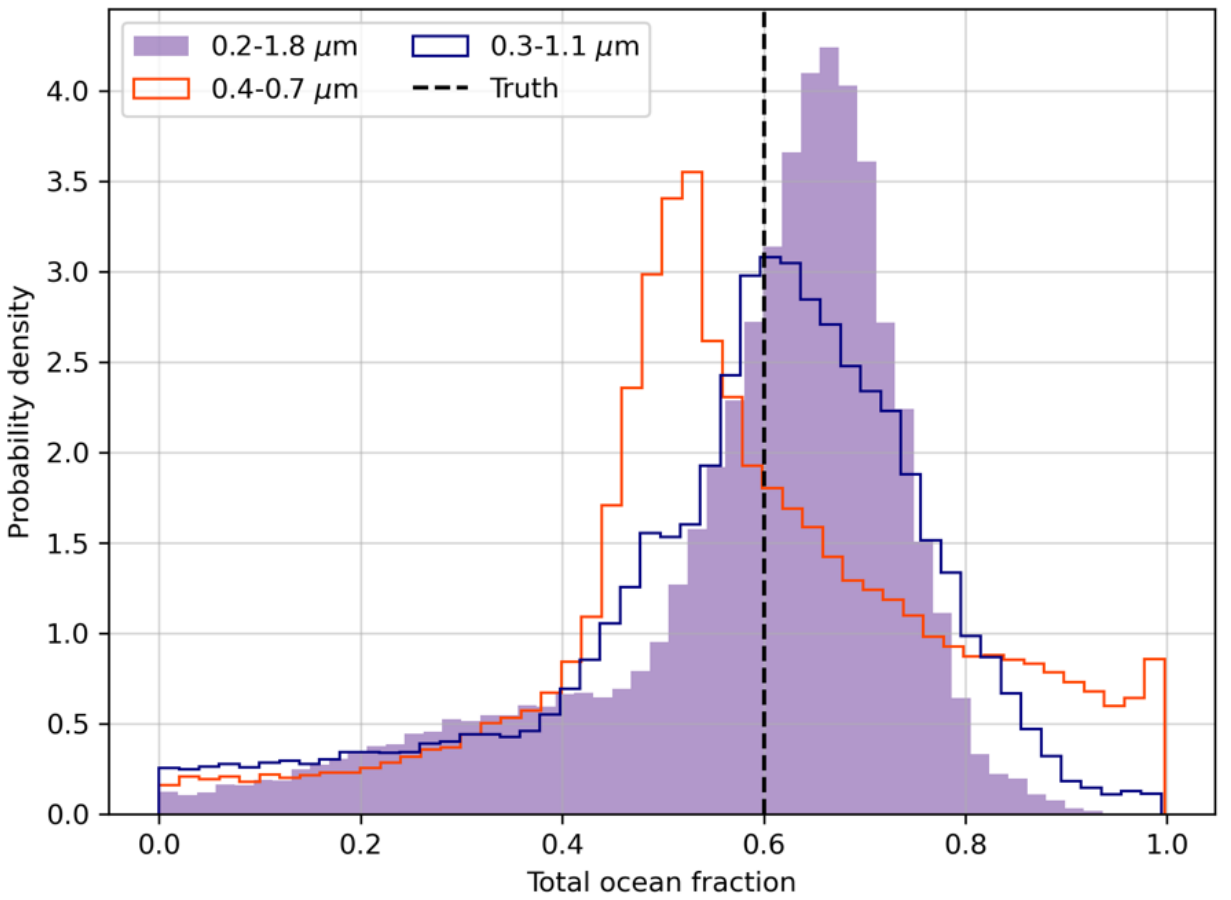

**Figure C1.** Ocean fraction posteriors for a selection of wavelength cutoffs at an SNR of 20 for our lifeless Earth scenario (60% ocean). Our minimum wavelength coverage for land detection, 0.3–1.1 $\mu$m in blue, provides a decent constraint on the ocean fraction, but all shown posteriors have extended tails toward zero ocean fraction.

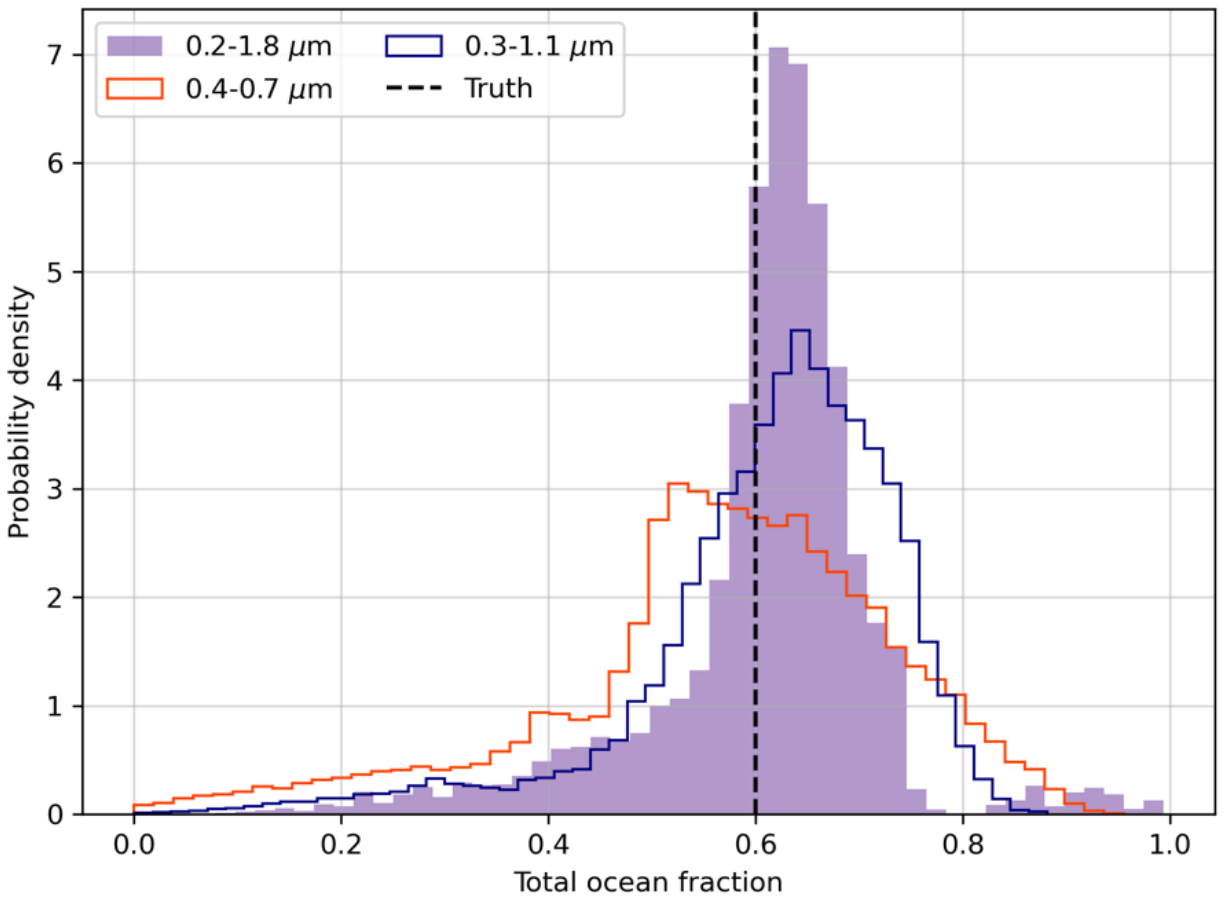

**Figure C2.** Ocean fraction posteriors for a selection of wavelength cutoffs at an SNR of 40 for our lifeless Earth scenario (60% ocean). At this higher SNR, the ocean detection is less affected by wavelength cutoff and provides a decent constraint for even 0.4–0.7 $\mu$m in orange. The extended tails of the distributions toward zero ocean are less pronounced compared to the SNR 20 case above.

## ORCID iDs

Anna Grace Ulses https://orcid.org/0000-0002-9031-0824
Joshua Krissansen-Totton https://orcid.org/0000-0001-6878-4866
Tyler D. Robinson https://orcid.org/0000-0002-3196-414X
Victoria Meadows https://orcid.org/0000-0002-1386-1710
David C. Catling https://orcid.org/0000-0001-5646-120X
Jonathan J. Fortney https://orcid.org/0000-0002-9843-4354